# Epitaxial growth of Ruddlesden-Popper neodymium nickelates $Nd_{n+1}Ni_nO_{3n+1}$ ($n$ = 1–5)

Wenjie Sun,[1,2] Yueying Li,[1,2] Xiangbin Cai,[3] Jiangfeng Yang,[1,2] Wei Guo,[1,2] Zhengbin Gu,[1,2] Ye Zhu[4,a)] and Yuefeng Nie[1,2,b)]

[1] National Laboratory of Solid State Microstructures, Jiangsu Key Laboratory of Artificial Functional Materials, College of Engineering and Applied Sciences, Nanjing University, Nanjing 210093, China

[2] Collaborative Innovation Center of Advanced Microstructures, Nanjing University, Nanjing 210093, China

[3] Department of Physics, The Hong Kong University of Science and Technology, Hong Kong, China

[4] Department of Applied Physics, Research Institute for Smart Energy, The Hong Kong Polytechnic University, Hong Kong, China

## ABSTRACT

A series of Ruddlesden-Popper nickelates, $Nd_{n+1}Ni_nO_{3n+1}$ ($n$ = 1–5), have been stabilized in thin film form using reactive molecular-beam epitaxy. High crystalline quality has been verified by X-ray diffraction and scanning transmission electron microscopy. X-ray photoelectron spectroscopy indicates the $n$-dependent valence states of nickel in these compounds. Metal-insulator transitions show clear $n$ dependence for intermediate members ($n$ = 3–5), and the low-temperature resistivities of which show logarithmic dependence, resembling the Kondo-scattering as observed in the parent compounds of superconducting infinite-layer nickelates.

Layered nickelates with chemical formula $R_{n+1}Ni_nO_{3n+1}$ ($R$ = trivalent rare-earth elements), which structurally belongs to Ruddlesden-Popper (RP) series[1], have long been considered as a promising candidate for realizing high temperature (high-$T_c$) superconductivity[2,3]. Indeed, some key features have been revealed in such compounds utilizing angle-resolved photoemission spectroscopy and X-ray absorption spectroscopy, such as pseudogap in $Nd_{2-x}Sr_xNiO_4$[4], large hole Fermi surface with $d_{x^2-y^2}$ character in $La_4Ni_3O_{10}$[5], and large orbital polarization in metallic $Pr_4Ni_3O_8$[6], etc. More recently, superconductivity has been discovered in hole-doped infinite-layer $NdNiO_2$[7-9] ($n$ = ∞ member of reduced RP series), and later in $PrNiO_2$[10] and $LaNiO_2$[11,12], which further triggers intensive research

---

interest in this field. Moreover, the required Ni 3$d$ orbital occupancy (or doping level) within the superconducting dome of hole-doped NdNiO$_2$[8,9] could be achieved by directly reducing RP phases with higher $n$ values ($n$ = 4, 5) without any cation substitution[6,13-15]. Thus, it may be possible to realize superconductivity in such compounds and potentially higher $T_c$ given enhanced superexchange interactions as already predicted[16]. Also, perovskite nickelates ($n = \infty$ member of RP series) possess varieties of exotic properties, among which the metal-insulator transition[17] (MIT) has attracted much attention, considering its potential compatibility for electronic devices. It is widely believed that the MIT in nickelates is mainly related to Ni-O-Ni bond angle[17,18], and can be modulated by structural engineering, such as changing $R$ size[19] or epitaxial strain[20]. As such, the MIT would be also modulated in RP series, since the insertion of extra $R$O layer into the perovskite framework along $c$ axis causes symmetry breaking in such phases. From these perspectives, RP nickelates with high $n$ value (3 < n < $\infty$) provide a new platform for in-depth study of the nickelate physics. However, they can only be obtained via layer-by-layer engineering[13], owing to thermodynamic instability. As a consequence, the investigation on their physical properties is still lacking up to now.

In this work, we report the epitaxial stabilization of Nd$_{n+1}$Ni$_n$O$_{3n+1}$ ($n$ = 1–5) thin films on (001) LaAlO$_3$ (LAO) substrates by reactive molecular-beam epitaxy (MBE). High crystalline quality is confirmed by the combination of high-resolution X-ray diffraction (XRD), atomic force microscopy (AFM), and scanning transmission electron microscopy (STEM). Transport measurements reveal the modulation of MIT by changing $n$ values, and the valence states of Ni across RP series are investigated by X-ray photoelectron spectroscopy (XPS).

A series of RP films were grown on (001)-oriented LAO single crystal substrates by MBE using DCA R450 system, and the growth process was monitored by an *in-situ* reflection high-energy electron diffraction (RHEED) system. An AlO$_2$-teminated atomically flat surface of LAO was obtained by the standard etching technique[21]. During the growth, the substrate temperature was fixed at 600 °C (except for $n$ = 1 member at 750 °C) and the background pressure was kept around 1 × 10$^{-5}$ torr of distilled ozone. All RP films were obtained by a shuttered layer-by-layer deposition, with the deposition sequence suggested in previous work[22,23]. The deposition time of NdO or NiO$_2$ monolayer has been precisely calibrated and corresponds to a single period of RHEED intensity oscillation [Fig. 1(a)] in the growth of NdNiO$_3$ using the co-deposition method[24]. It should be noted that the slight overall intensity drift is caused by the background pressure oscillations, rather than the wrong dosage of individual sources, which is frequently seen in SrTiO$_3$ growth[24]. All films were grown on the same day to minimize the possible variation in beam fluxes, background pressure, etc. The inset in Fig. 1(a) shows the RHEED pattern of Nd$_6$Ni$_5$O$_{16}$ ($n$ = 5 member of RP series) using the shuttered growth method, and the sharp and bright diffraction spots indicate its high surface crystallinity.

The crystalline quality of RP films was further examined using XRD. As shown in Fig. 1(b), sharp diffraction peaks in $2\theta$-$\omega$ scans suggest single phase of present RP series. Also, there is no obvious peak splitting, implying the right periodic sequence during growth, since the minor deviation (1 %) of monolayer dosage will lead to peak splitting and shift in RP films[25]. Reciprocal space mapping (RSM) indicates RP films are fully constrained to the substrate, with a representative $n = 5$ member shown in Fig. 1(c). The out-of-plane lattice constants calculated from corresponding RSM images are 12.43 ± 0.01 Å, 20.17 ± 0.02 Å, 27.76 ± 0.01 Å, 35.52 ± 0.02 Å, 43.14 ± 0.01 Å, 3.85 ± 0.01 Å for the $n = 1$–5 and $\infty$ phases, respectively. Fig. 1(d) shows the AFM topography of the $n = 5$ film, in which the smooth surface with step height of ~0.38 nm (as shown in the inset) reflects the layer-by-layer growth mode as expected.

It has been documented that some unexpected defects, such as intergrowth[25] and SrO double-layer missing[22], will appear during the growth of $Sr_{n+1}Ti_nO_{3n+1}$ RP series. Such defects will prohibit the identification of intrinsic properties, for example dielectric properties[26,27]. As such, atomic-resolution STEM was performed for the $n = 5$ film to examine the local nanostructures. As shown in Fig. 2(a), cross-sectional annular dark-field (ADF) STEM image and the corresponding electron diffraction pattern reveal the layered structure consisting of alternatively stacked perovskite units and NdO-NdO rock-salt double layers, as expected. Moreover, sharp interfaces with no obvious ionic intermixing can be verified through atomic-resolution energy-dispersive X-ray spectroscopy (EDX) maps [Figs. 2(c)-(e)]. All prove the feasibility of non-stoichiometric deposition method[22] in growing RP nickelates with any $n$ value at the atomic level.

A prominent feature of layered nickelates is the ability to engineer $3d$ orbital occupancy of Ni between $3d^8$ (divalent for $n = 1$) and $3d^7$ (trivalent for $n = \infty$), without chemical substitutions. In order to verify this feature and quantify the evolution of Ni valence states, we performed XPS measurements and the results are summarized in Fig. 3. All XPS data were collected at room temperature in an ultra-high-vacuum chamber using Al K$\alpha$ radiation ($h\nu$ = 1486.6 eV), and calibrated by C 1s peak position (284.8 eV). It has been reported that multiplet splitting is necessary for verifying chemical states of Ni, due to contributions from electronic coupling and screening effect[28]. Thus, we apply $Ni^{3+}$ and $Ni^{2+}$ multiplet envelope from Ref. 28 to fit the Ni $2p_{3/2}$ spectra of $NdNiO_3$ and $Nd_2NiO_4$ respectively, using XPST package[29]. The results are quite satisfying [Fig. 3(a)], indicating the expected dominant valence state of Ni being 3+ for $NdNiO_3$ and 2+ for $Nd_2NiO_4$. The core-level peak locates at around 854.4 eV for $NdNiO_3$ and 853.5 eV for $Nd_2NiO_4$, consistent with the results for $SmNiO_3$ ($Ni^{3+}$, 853.9 eV)[30] and NiO ($Ni^{2+}$, 853.4 eV)[31]. However, the fitting of intermediate valence (between $Ni^{2+}$ and $Ni^{3+}$) spectra considering multiplet splitting would be much complex, for two sets of multiplet envelope needed to be taken into account[28], which would complicate the refinement of fitting parameters and reduce the reproducibility of fitting results[32]. Thus, here we use non-linear least-squares fitting (NLLSF) method[32] to fit the Ni $2p_{3/2}$ spectra of other members ($n$ = 2–5) using Avantage software. As shown in Fig. 3(b), the fitted curves of $NdNiO_3$ and $Nd_2NiO_4$ were used as reference spectra to determine the $Ni^{3+}/Ni^{2+}$ ratio, which was calculated from the area under each reference. The valence states of Ni determined by corresponding $Ni^{3+}$/Ni ratios show good agreement with nominal values [Fig. 3(c)], which were calculated from the chemical formula of each compound if considering Nd always being trivalent. Indeed, the main peak positions of Nd $3d_{5/2}$ spectra are nearly the same across RP series (not shown). Thus, RP structures can effectively adjust the $d$ orbital occupancy of Ni, providing an opportunity to achieve similar electronic structure as high-$T_c$ cuprates.

The temperature-dependent resistivities of RP films were measured using a standard collinear four-point probe method[33], as shown in Fig. 4(a). Gold electrodes with equal spacing of 1 mm and thickness of 70 nm were deposited by ion sputtering to achieve ohmic contacts. As one of the most fascinating properties among perovskite rare-earth nickelates, MIT is considered most likely originating from the bond disproportionation[17,18]. In addition, it can be tuned by various manners, such as strain engineering[20], dimensionality control[34], and electrolyte gating[35], etc. As shown in Fig. 4(b), the $NdNiO_3$ film shows sharp MIT plus thermal hysteresis behavior, and the resistivity change reaches nearly 3 orders of magnitude across MIT, in line with previous reports[35,36] and also its single-crystalline $PrNiO_3$ counterpart[37,38]. This suggests the near optimal stoichiometry and high crystalline quality of present $NdNiO_3$ film, due to the fact that cation off-stoichiometry will dramatically change both the onset temperature and the magnitude of resistivity change at the MIT[39,40]. Furthermore, this in turn guarantees the right deposition time for each atomic layer of RP film growth, as they are determined from co-deposition process of $NdNiO_3$. For $n$ = 1 and 2 films, insulating behavior is persistent in the whole temperature range (4–300 K), which most likely originates from charge ordering[4,41]. Interestingly, MIT and thermal hysteresis behavior appear for the members with $n \geq 3$, although the magnitude of resistivity change across MIT and thermal hysteresis are much smaller compared to those of $NdNiO_3$. And this is very similar to the case of some bulk RP nickelates, in which the hysteresis behavior is also very weak or even eliminated, perhaps manifesting the second-order or weakly first-order nature of this transition in $n \geq 3$ RP phases[42-44].

Also, the transition temperature ($T_{MI}$) exhibits clear $n$ dependence, decreasing with increasing $n$ value, which is clearly demonstrated in Fig. 4(c). And this may relate to the increasing of electron hopping strength caused by the weakening of dimensionality effect as $n$ increases[13], akin to the case in $R$NiO$_3$[17]. Enhanced orbital overlaps between Ni 3$d$ and O 2$p$ from LuNiO$_3$ to LaNiO$_3$ leads to higher electron hopping strength, as a consequence, $T_{MI}$ decreasing from 600 K (in LuNiO$_3$) to 0 K (in LaNiO$_3$)[17]. In addition, the increasing of electron hopping strength would also be responsible for the decreasing of resistivity at the metallic phase as $n$ changes from 1 to ∞. Specially, for $n$ = 3 member, both MIT and metal-to-metal transition around 160 K have been documented for some polycrystalline counterparts[42-44], but the origins are still in controversy, which may be associated with the oxygen off-stoichiometry or phase purity[43]. In our case, all films are post-annealed in 1.5 bar pure oxygen at 350 °C for 10 hours before transport measurements to minimize the effect of oxygen vacancies, and the MIT behavior is robust within a series of RP members with high $n$ values. Fig. 4(d) shows the semi-logarithmic resistivity curves of $n$ = 3–5 RP films. Clear ln $T$ dependence at low temperatures may indicate the Kondo scattering as the origin of resistivity upturn. Similar logarithmic dependence has been observed in Nd$_4$Ni$_3$O$_{10-\delta}$ polycrystalline sample[44] and parent compounds of nickelate superconductors[7,45] (NdNiO$_2$ and LaNiO$_2$), in which the Ni-moment is regarded as a Kondo-scattering center[46]. Thus, similar low-temperature Kondo scattering may happen as well in the RP phases, making them promising candidates to reproduce Ni-based superconductivity if Kondo physics plays an important role here as theoretically predicted[46].

In conclusion, a series of Nd-based RP nickelates ($n$ = 1–5) have been synthesized using MBE, especially the metastable members with high $n$ values. For Nd$_6$Ni$_5$O$_{16}$, both the square-planar crystal structure and 3$d^{8.8}$ orbital occupation will be satisfied, upon further topochemical reduction[6,15]. However, the exploration of reduction parameters is still challenging, and lies beyond the scope of current work. Transport measurements demonstrate effective modulation of $T_{MI}$ by changing $n$ values. Also, low-temperature resistivities of RP members with high $n$ values show a clear logarithmic dependence, similar to that of NdNiO$_2$ and LaNiO$_2$, suggesting the possible occurrence of Kondo scattering. Moreover, the valence states of Ni show good correspondence with the nominal values in present RP series, as indicated by XPS. Our work thus demonstrates the effectiveness of layer-by-layer growth method for stabilizing the metastable RP phases and provides another appealing platform for exploring high-$T_c$ superconductivity and exotic properties in layered nickelates.

The authors thank Jierong Wang and Haoying Sun for proofreading the manuscript. This work was supported by the National Natural Science Foundation of China (Grant Nos. 11774153, 11861161004, and 51772143), the Fundamental Research Funds for the Central Universities (Grant No. 0213-14380198, 0213-14380167), the Hong Kong Research Grants Council (RGC) through the NSFC-RGC Joint Research Scheme (Grant No. N_PolyU531/18), and the Hong Kong Polytechnic University grant (ZVGH).

The data that support the findings of this study are available from the corresponding author upon reasonable request.

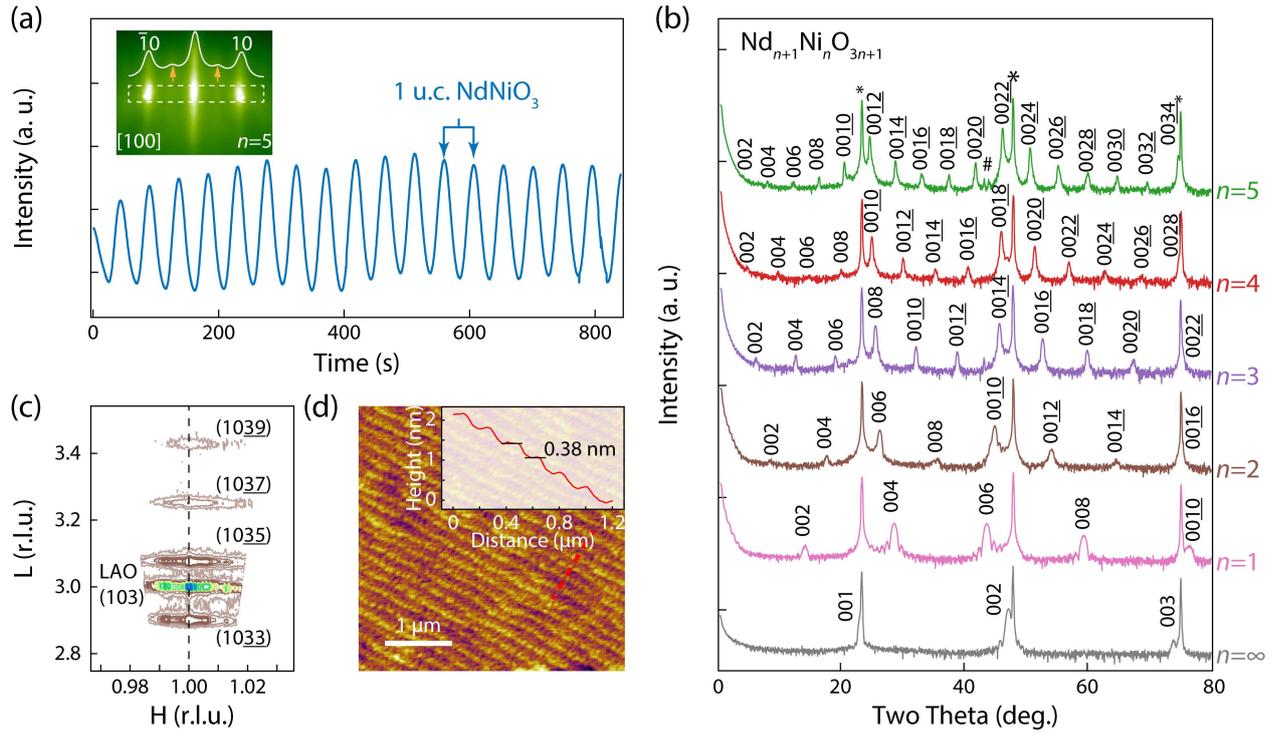

FIG. 1. (**a**) RHEED intensity oscillations during the co-deposition growth of NdNiO$_3$ films. Inset shows the RHEED pattern of the Nd$_6$Ni$_5$O$_{16}$ film taken along [100] pseudocubic direction. White solid line represents the intensity profile integrated along the vertical axis of the rectangular region marked with white dashed line. The half order peaks caused by oxygen octahedral rotations were indicated by yellow arrows. (**b**) 2$\theta$-$\omega$ XRD scans of 9-u.c.-thick Nd$_{n+1}$Ni$_n$O$_{3n+1}$ RP films with $n$ = 1–5 and 20-u.c.-thick NdNiO$_3$ film ($n$ = ∞ member of RP phases). Diffraction peaks of LAO substrate are denoted by asterisks. All plots are vertically offset for clarity. The peak marked by # are background diffractions appearing occasionally depending on the exact configuration of the instrument. (**c**) RSM image of the 9-u.c.-thick Nd$_6$Ni$_5$O$_{16}$ around LAO (103). (**d**) AFM image of the 9-u.c.-thick Nd$_6$Ni$_5$O$_{16}$ and corresponding line cut (inset) along the red dashed line. The root-mean-square roughness is 94 pm.

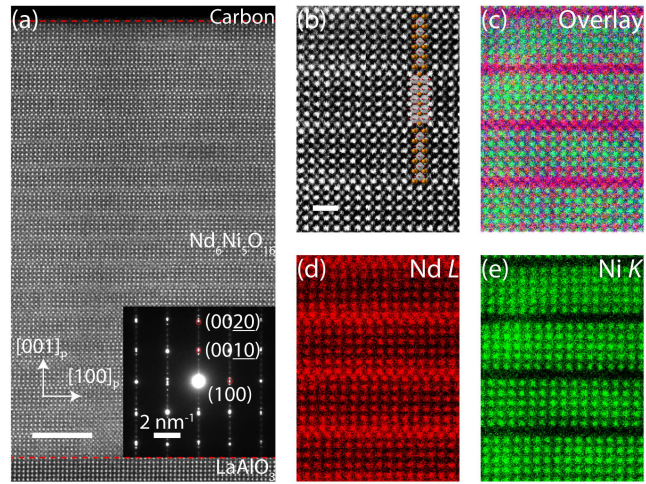

FIG. 2. (**a**) Cross-sectional ADF-STEM image of a 9-u.c.-thick Nd$_6$Ni$_5$O$_{16}$ ($n$ = 5) film and the corresponding electron diffraction pattern shown in the inset. The subscript p denotes the pseudocubic notation. Scale bar, 5 nm. (**b**) Enlarged view of the ADF-STEM image shown in (**a**) with the crystal structure model overlaid. Scale bar, 1 nm. (**c-e**) Atomic-resolution EDX maps of Nd + Ni, Nd, and Ni acquired from the same area as (**b**).

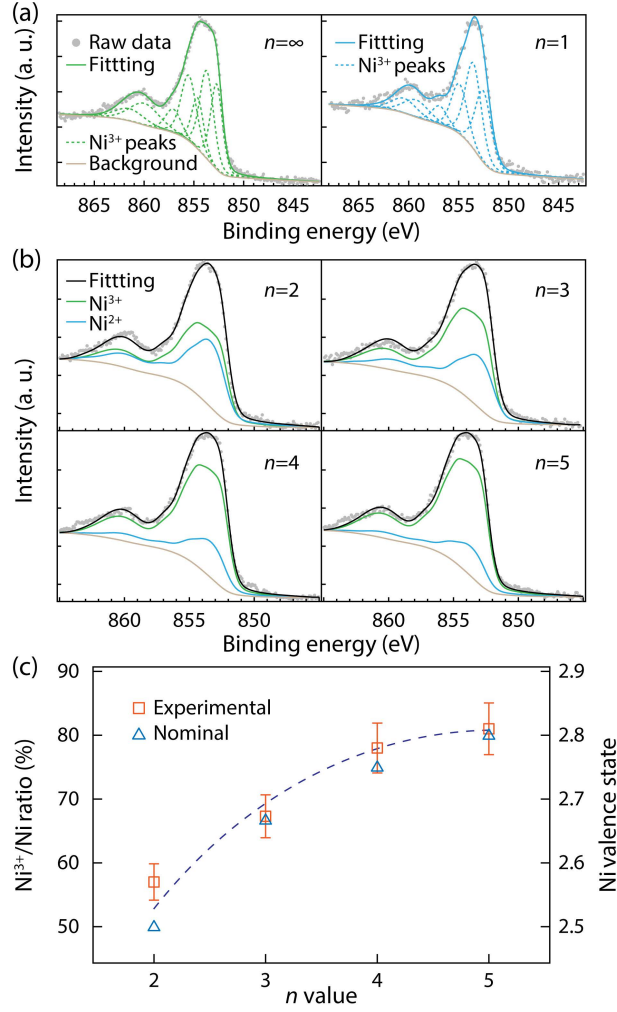

FIG. 3. (**a**) The Ni $2p_{3/2}$ spectra of NdNiO$_3$ ($n = \infty$) and Nd$_2$NiO$_4$ fitted by Ni$^{3+}$ and Ni$^{2+}$ multiplet splitting peaks, respectively. (**b**) The Ni $2p_{3/2}$ spectra of $n = 2$–5 RP films fitted using NNLSF method as mentioned in the main text. (**c**) The proportion of Ni$^{3+}$ and corresponding Ni valence states in $n = 2$–5 RP films estimated from NNLSF analysis. Error bars represent uncertainty of the fitting process. Dashed line is the guide to the eyes.

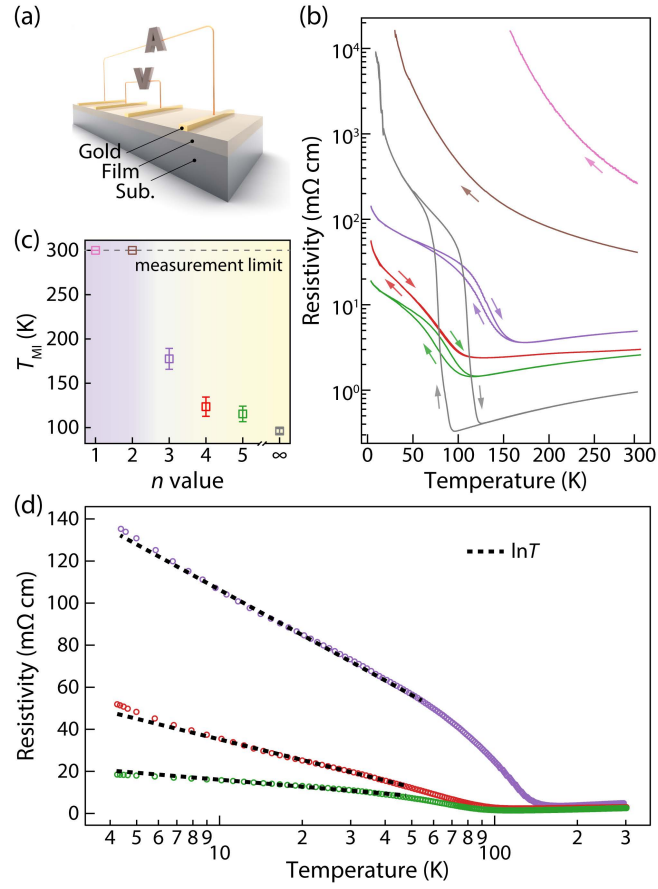

FIG. 4. (**a**) Schematic of the collinear four-point probe geometry used for the transport measurements. (**b**) Temperature-dependent resistivities of $n$ = 1–5 RP films and 22-u.c.-thick $NdNiO_3$ film. Arrows indicate the temperature-sweeping direction. The temperature-sweeping rate is kept at 2 K/min. The color legend is the same as the data series in Fig. 1(b). (**c**) $T_{MI}$ of RP films during the cooling process. (**d**) Semi-logarithmic plot of resistivities of $n$ = 3–5 RP films during the cooling process. The black dashed lines are the corresponding $\ln T$ fits. The color legend is the same as the data series in Fig. 1(b).